# Revealing the role of nitrogen on hydride nucleation and stability in pure niobium using first principles calculations

P. Garg[1], S. Balachandran[2], I. Adlakha[1,3], P. J. Lee[2], T. R. Bieler[4], and K. N. Solanki[1*]

[1]School for Engineering of Matter, Transport, and Energy, Arizona State University, Tempe, AZ 85287, USA

[2]The Applied Superconductivity Center, National High Magnetic Field Laboratory, Florida State University, Tallahassee FL 32310, USA

[3]Department of Applied Mechanics, Indian Institute of Technology-Madras, Chennai, India

[4]Department of Material Sciences and Engineering, Michigan State University, East Lansing, MI, USA

*Corresponding Author: (480) 965-1869; (480) 727-9321 (fax), E-mail: kiran.solanki@asu.edu

**Abstract**

Niobium provides the basis for all superconducting radio frequency (SRF) cavities in use, however, hydrogen is readily absorbed by niobium during cavity fabrication and subsequent niobium hydride precipitation when cooled to cryogenic temperatures degrades its superconducting properties. In the last few years the addition of dopant elements such as nitrogen has been experimentally shown to significantly improve the quality factor of niobium SRF cavities. One of the contributors to Q degradation can be presence of hydrides; however, the underlying mechanisms associated with the kinetics of hydrogen and the thermodynamic stability of hydride precipitates in the presence of dopants are not well known. Using first principles calculations, the effects of nitrogen on the energetic preference for hydrogen to occupy interstitial sites and hydride stability are examined. In particular, the presence of nitrogen significantly increased the energy barrier for hydrogen diffusion from one tetrahedral site to another interstitial site. Furthermore, the beta niobium hydride precipitate became energetically unstable upon addition of nitrogen in the niobium matrix. Through electronic density of states and valence charge transfer calculations, nitrogen showed a strong tendency to accumulate charge around itself, thereby decreasing the strength of covalent bonds between niobium and hydrogen atoms leading to a very unstable state for hydrogen and hydrides. These calculations show that the presence of nitrogen during processing plays a critical role in controlling hydride precipitation and subsequent SRF properties.

# 1. Introduction

Niobium is the primary material for manufacturing superconducting radio-frequency (SRF) cavities for high-performance particle accelerators because it has both the highest lower magnetic field ($H_{c1}$~190 mT) and highest superconducting transition temperature ($T_c$ = 9.25 K) of any element along with high ductility that enables forming of complex cavity shapes [1,2]. However, niobium readily absorbs hydrogen during chemical processing steps, and these absorbed hydrogen atoms segregate near defects (point, line or planar defects) or precipitate into hydride phases [3–5]. High concentrations of hydrogen on niobium surfaces degrade the properties of niobium desired for SRF applications [6,7]. For instance, segregation of interstitial hydrogen atoms and hydride precipitation leads to localized breakdown of surface superconductivity due to proximity coupling during cavity operation at radio frequencies [8–12]. Advances in cavity preparation techniques such as high temperature heat treatments (>600°C) followed by low temperature baking (120°C), diffusion of nitrogen into sub-surface layers of Nb have been developed to improve the cryogenic efficiency (operation at higher accelerating fields) of current and future accelerators [10,13–16]. For instance, recent experiments show significantly higher intrinsic quality factors ($Q_0$), for cavity surfaces treated in a nitrogen atmosphere as compared to surfaces with no nitrogen treatment [17–21]. This intrinsic quality factor is a measure of cavity performance that quantifies energy loss in the context of storing a large amount of energy [22]. Furthermore, the effect of diffusion of nitrogen on the radio frequency surface resistance and the trapped magnetic flux sensitivity

(energy loss mechanisms) has been reported to depend strongly on cavity processing techniques [23,24]. However, many fundamental questions related to the kinetic and thermodynamic aspects of such improvements remain elusive.

In general, the observed improvement in performance of nitrogen doped cavities can be attributed to several parameters, such as reduction of the surface resistance [22,25] and trapping of interstitial hydrogen by nitrogen that prevents hydrogen clustering or hydride nucleation [25,26]. Hydride formation leads to plastic distortions in the niobium lattice, which helps to easily observe the [13] hydride pits ex-situ due to strain contrast by back scatter electron imaging under a microscope. Hydride formation in niobium wires that underwent cavity relevant heat treatments, including a traditional 800°C/3 h heat treatment (control), versus nitrogen doping/diffusion heat treatment was compared. The nitrogen treatment involved, 800°C/3h heat treatment in vacuum followed by a two minute flow of $N_2$ gas at a partial pressure of 5 mTorr after which the $N_2$ supply was shut, and the sample was soaked at 800°C for six minutes. This recipe is commonly referred to as "8002N6" in the SRF community [20]. This recipe is currently being used in the production of cavities for the LCLS-II light source upgrade at SLAC. The 8002N6 recipe with nitrogen diffusion through the surface of the wire show low hydride pit concentration in regions where nitrogen concentration is higher (Figure 1a and 1b) [27]. Within the first 50μm from the surface, a 60% decrease in the density of hydrides was observed in the niobium coupons treated in a nitrogen atmosphere compared to the control sample (Figure 1c). These observations provide the opportunity to seek a better mechanistic understanding of how nitrogen doping influences the kinetic stability of hydrogen and the thermodynamic stability of hydride precipitates in niobium.

Therefore, the main goal of the present work is to understand the mechanisms associated with interstitial nitrogen acting as a trapping site for hydrogen and how it inhibits hydride precipitation in niobium as observed in experiments. Previous first-principles calculations confirmed that hydrogen absorption is energetically favorable at interstitial tetrahedral sites in niobium [2,28]. Thus, first-principles calculations are used to examine the effect of nitrogen on hydrogen absorption and the stability of hydride precipitates in niobium.

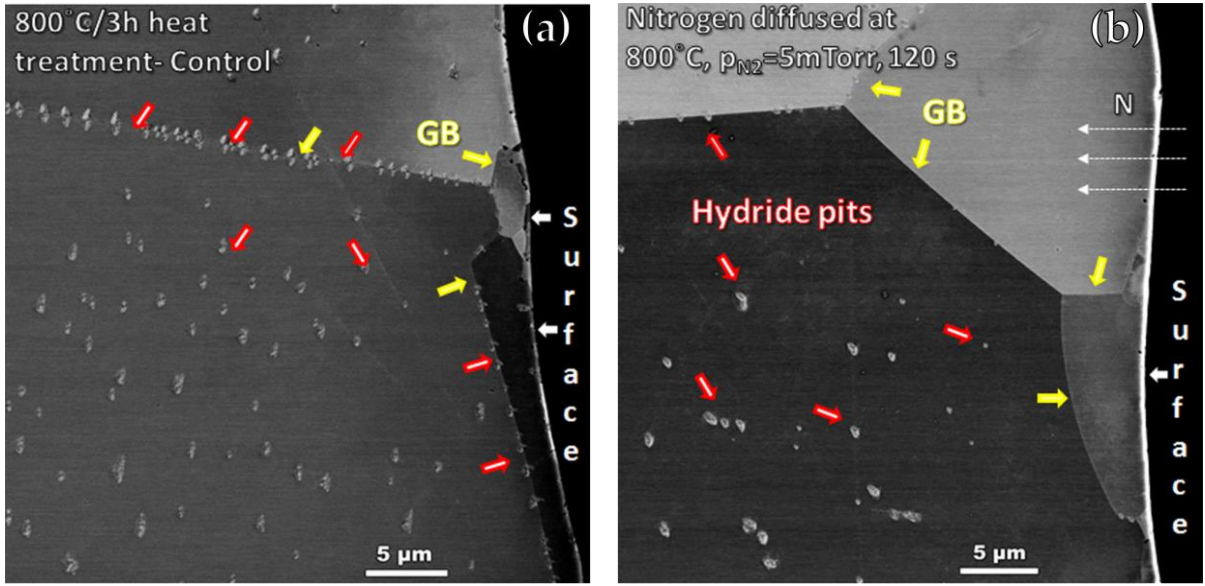

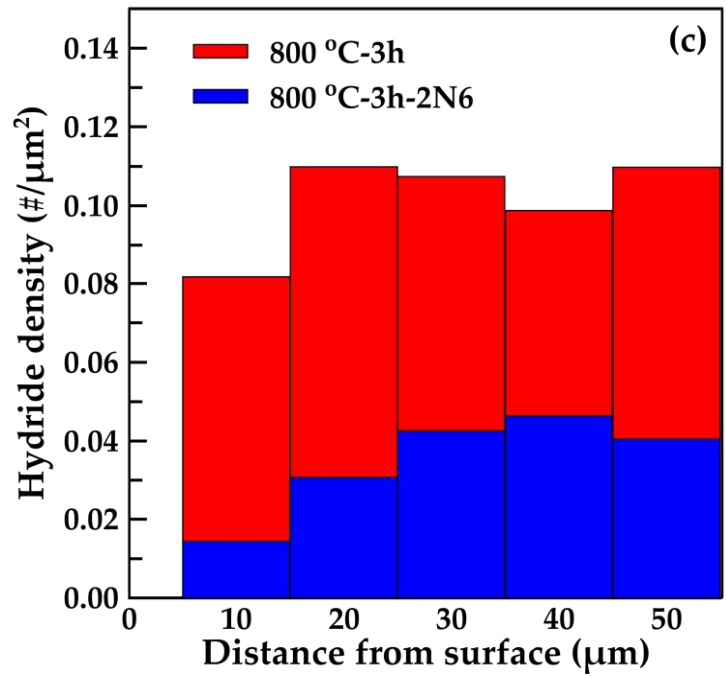

**Figure 1:** Microstructure of niobium coupon samples treated (a) without nitrogen atmosphere and (b) in nitrogen atmosphere show low hydride concentration for the nitrogen treated niobium surfaces. (c) The hydride density measured from the surface is nearly constant in the 800°C-3h (control) sample, and lower in the sample where surface N concentration is higher even up to distances of 50μm.

## 2. Computational details

First-principles calculations were performed using the Vienna Ab-initio Simulation Package (VASP) based on the density functional theory (DFT) [29,30]. The projector augmented wave (PAW) pseudopotentials[31,32] using the Perdew-Burke-Ernzerhof (PBE) [33] exchange-correlation formulation were used to represent the nuclei with valence electrons. A plane wave basis set with an energy cutoff of 550 eV, 350 eV and 550 eV were used for niobium, hydrogen and nitrogen atoms, respectively. The Monkhorst-Pack K-point mesh of 12 x 12 x 12 was employed to carry out the Brillouin-zone integrations [34] and the ions were relaxed with a force and energy convergence criteria of 0.01 eVÅ$^{-1}$ and 10$^{-6}$ eV respectively. The calculated niobium lattice parameter of 3.31 Å was found to be in good agreement with the reported experimental and theoretical values [2,6].

## 3. Results and discussions

*3.1 Suppression effect of N doping on binding of H in Nb*

Absorption of hydrogen and nitrogen is energetically favorable at interstitial tetrahedral (T) and octahedral (O) sites in niobium (-0.28 eV and -1.98 eV), respectively. A 2 x 2 x 2 body centered cubic niobium supercell (16 niobium atoms) was used to examine the binding behavior one hydrogen and one nitrogen atom within the niobium lattice, i.e. a concentration of 1/16. The binding energy (ΔE) for hydrogen at different interstitial sites in niobium with a nitrogen atom located at an octahedral site was calculated using the following equation (see [35–37]):

$$\Delta E = E_{Nb_{16}N_1H_1} - (E_{Nb_{16}N_1} + 0.5 * E_{H_2}) \qquad (1)$$

where $E_{Nb_{16}N_1H_1}$ represents the energy of the niobium lattice with nitrogen and hydrogen atoms present at their respective interstitial sites, $E_{Nb_{16}N_1}$ represents the energy of niobium lattice with a nitrogen atom doped at an octahedral site and $E_{H_2}$ represents the energy of hydrogen in gas-phase. Negative values of the $\Delta E$ implies that hydrogen binding is energetically favorable, while positive values indicate that it is favorable for hydrogen to leave the interstitial site and go into another favorable site.

For the case of hydrogen doped at different tetrahedral sites near nitrogen, it was found that hydrogen binding was no longer energetically favorable at sites T1 and T2, identified in Figure 2. However, the binding energy gradually decreased and became negative for the interstitial sites located away from nitrogen, such as T3, T4 and O sites, as shown in Figure 2. Thus, nitrogen suppressed hydrogen binding in niobium but the level of suppression gradually diminished for hydrogen present at interstitial sites farther away from nitrogen, approximately 2.5 Å. Furthermore, the suppressing effect of nitrogen is dependent on the local concentration of both nitrogen and hydrogen present in niobium. Interestingly, both T4 and O interstitial sites away from nitrogen were found to have similar binding energies for hydrogen binding in niobium.

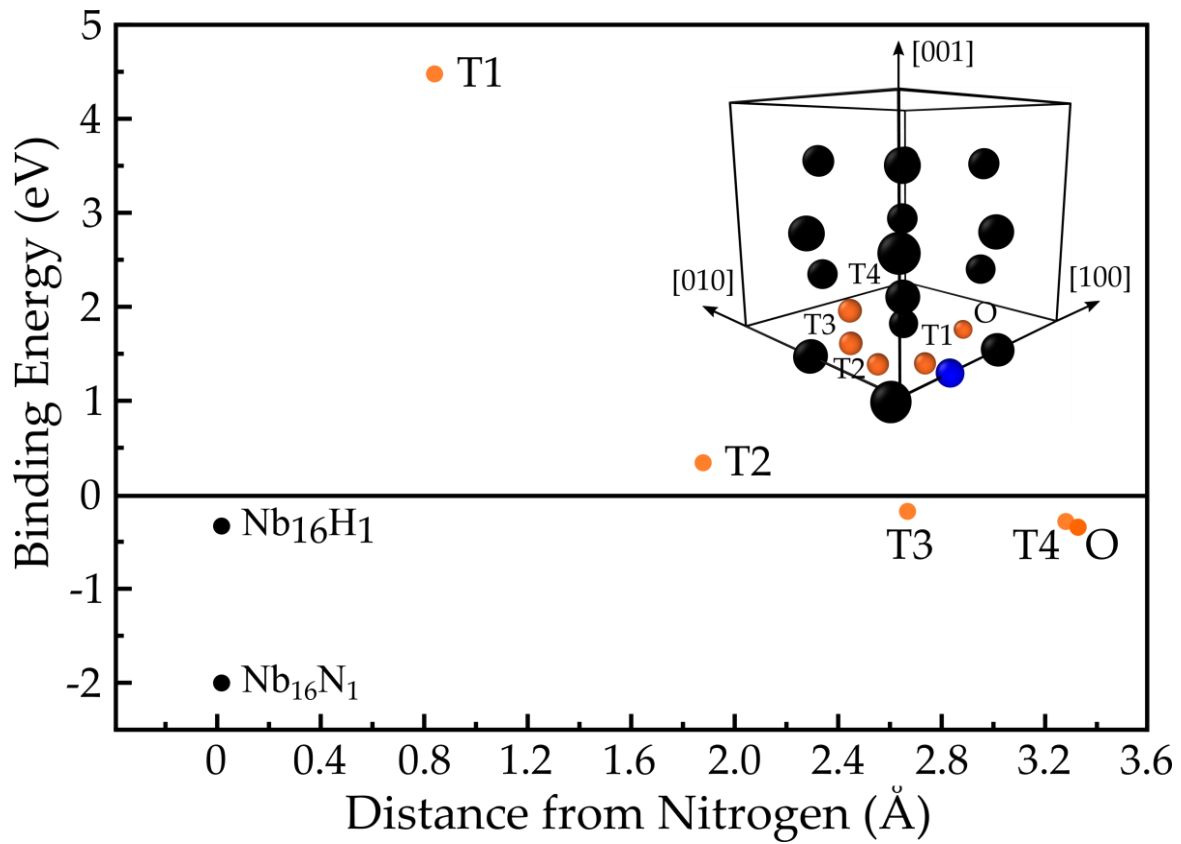

**Figure 2**: Plot of the binding energy of hydrogen at four different tetrahedral sites and one octahedral site near a nitrogen atom located at an octahedral site in niobium versus distance of interstitial sites away from the nitrogen atom. Niobium, hydrogen and nitrogen atoms are represented by dark grey, red and blue spheres, respectively.

To understand the effect of nitrogen on the binding energy of hydrogen in niobium, the electronic structure of bulk niobium with interstitially absorbed hydrogen and nitrogen atoms was examined using the electronic density of states (DOS) and valence charge transfer calculations. This provides insights into the changes in bonding between the interstitial and the host metal atoms. The DOS curve describes the number of states available to be occupied per interval of energy at each energy level [38]. To calculate the valence charge transfer resulting from the addition of different interstitial atoms, the ground state non-interacting valence charge densities were subtracted from the valence charge density of the interacting system [39,40]. Thus, the valence charge transfer (Δρ) was calculated as:

$$\Delta\rho = \rho_{Nb-H-N} - \rho_{Nb} - \rho_H - \rho_N \qquad (2)$$

where $\rho_{Nb-H-N}$ represents the valence charge density of interacting Nb-H-N system, and $\rho_{Nb}$, $\rho_H$ and $\rho_N$ represent the valence charge density of non-interacting and isolated niobium, hydrogen and nitrogen atoms, respectively. The VESTA (visualization for electronic and structural analysis) software package [41] was used to extract and visualize the valence charge transfer contours from the first-principles calculations.

First, the DOS curve of bulk niobium with interstitially absorbed hydrogen was examined, i.e., without nitrogen atoms. The DOS curve shows an overlap between the states of niobium and hydrogen at -7 eV indicating the formation of covalent bonds between niobium and hydrogen atoms see Figure S1a. A similar overlap characteristic of covalent bonding has been reported in previous studies for various metal hydride systems [42–44]. The covalent bonding was further verified by assessing the valence charge density around niobium and hydrogen atoms shown in Figure S1b. Here, the valence charge around black niobium atoms was attracted towards the orange hydrogen atom (from the cyan toward the yellow iso-charge surface), contributing to the energetic preference of forming a covalent bond. However, in the presence of nitrogen at an O site (blue atom that appears gray) and hydrogen at the T1 tetrahedral site, the states of nitrogen overlapped with the states of niobium and hydrogen (Figure 3a). As a result, the interactions between the states of niobium and hydrogen were interrupted by the presence of nitrogen, suggesting a decrease in the extent of covalent bonding between niobium and hydrogen atoms. Also, the development of repulsive interactions due to the accumulation of valence charge between hydrogen and nitrogen atoms (yellow iso-charge surface, Figure 3b) made the structure energetically unstable. This indicates a positive energy for hydrogen binding at the T1 site in bulk niobium with a nitrogen atom at an O site. Similar effects of nitrogen on covalent bonding between niobium and hydrogen were observed when hydrogen was present at the T2 tetrahedral site (Figure 3c). However, the repulsive interactions between nitrogen and hydrogen are weaker, indicated smaller binding energy, due to the increased distance between them at the T2 site (Figure 3d). When a hydrogen atom is located at a T4 or O site, states of niobium and hydrogen overlapped (Figure 3e), irrespective of the presence of nitrogen in bulk niobium. Thus, the interactions between niobium and hydrogen were not affected by nitrogen and the covalent bonding between niobium and hydrogen atoms was conserved when hydrogen atom was present away from nitrogen. Furthermore, the valence charge accumulated around the hydrogen and the nitrogen atom was attracted towards the charge depleted region (cyan iso-surface, Figure 3f-g) between them, contributing to the energetic preference of hydrogen to occupy T4 or O site in niobium.

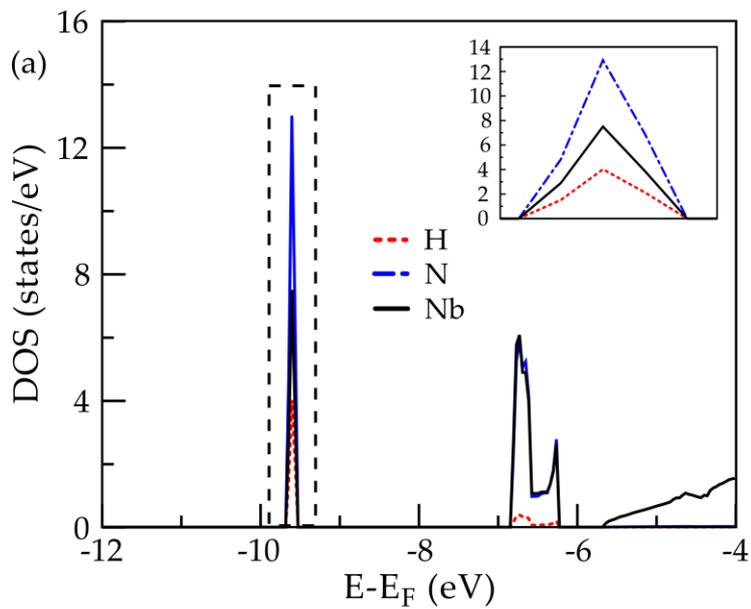
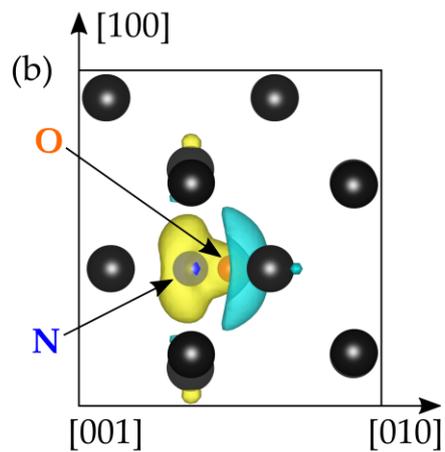
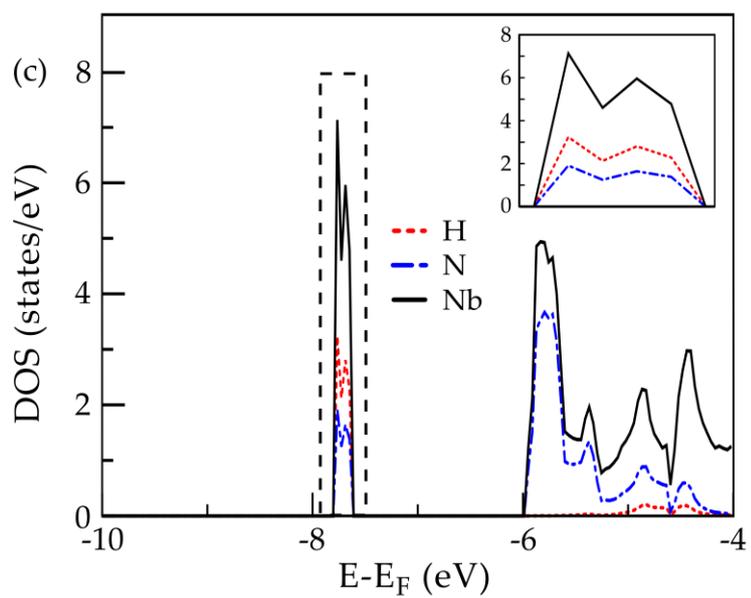
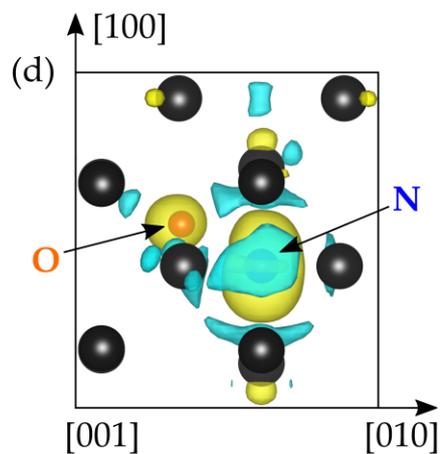
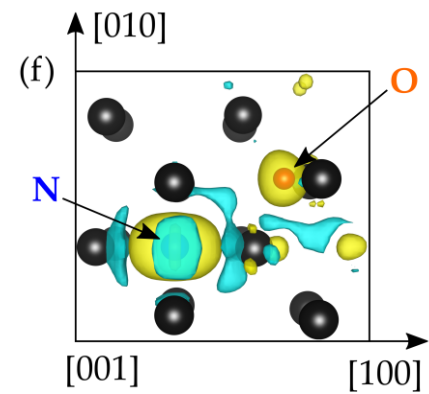
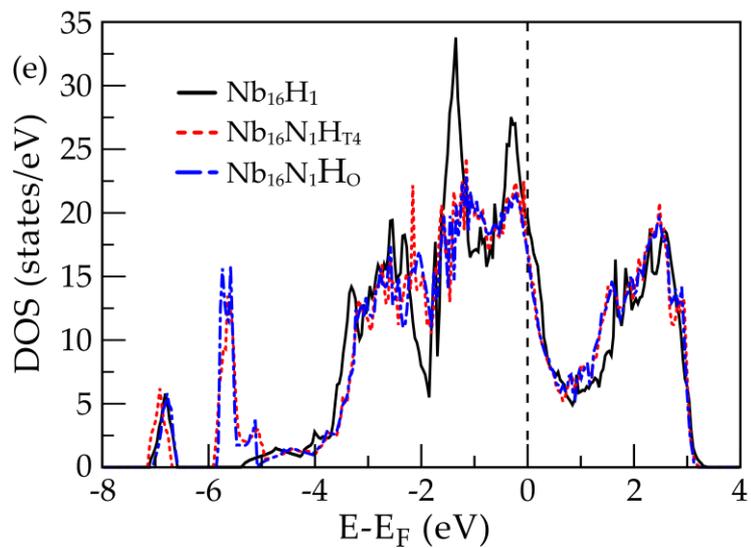
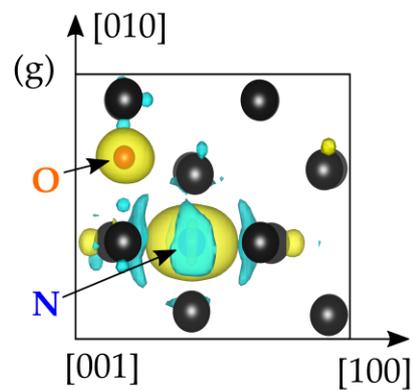

**Figure 3:** (a) DOS and (b) charge transfer of $Nb_{16}N_1$ with an H atom at the T1 site. (c) DOS and (d) charge transfer of $Nb_{16}N_1$ with H atom at T2 site. (e) DOS for $Nb_{16}H_1$ compared to the DOS of $Nb_{16}N_1$ with one H atom at the T4 site and for one H atom on the O site. (f, g) The valence charge transfer plots for $Nb_{16}N_1$ with H atom at T4 site and O sites, respectively. Yellow and cyan iso-charge surfaces represent $0.027e/Å^3$ of charge accumulation and depletion, respectively in (b), (d), (f) and (g). Niobium atoms are represented by black spheres in the valence charge transfer contours.

*3.2 Effect of N on H diffusion in Nb*

Next, the effect of nitrogen on the kinetic stability of hydrogen in the niobium lattice was investigated. It is critical to study the effect of nitrogen on the activation energy barrier for hydrogen diffusion as it affects the migration and retention of hydrogen in bulk niobium. To determine the activation barriers and the minimum energy pathways associated with hydrogen diffusion, the nudged elastic band (NEB) method [45] was used within the DFT framework. The NEB calculations were performed using the fast-inertial relaxation engine (FIRE) optimizer [46] with eight intermediate images, selected after reaching convergence based on the number of images. Other input parameters and convergence criteria were kept the same as for the ground state calculations. In the niobium lattice, the activation energy barrier for hydrogen diffusion from one tetrahedral site to the nearest tetrahedral site was found to be 0.18 eV (Figure 4a), which compares well with the reported literature value [47]. However, with the addition of nitrogen, the activation energy barrier along the minimum energy path for hydrogen diffusion in the niobium lattice increased significantly as shown in figure 4b. In the presence of nitrogen in niobium, a hydrogen atom will diffuse from one tetrahedral site to the 2nd nearest tetrahedral site, not the nearest tetrahedral site, via a metastable octahedral site with an activation energy barrier of 0.457 eV. Thus, nitrogen increases the energy barrier for hydrogen diffusion and hence, decreases the rate of hydrogen diffusion in bulk niobium.

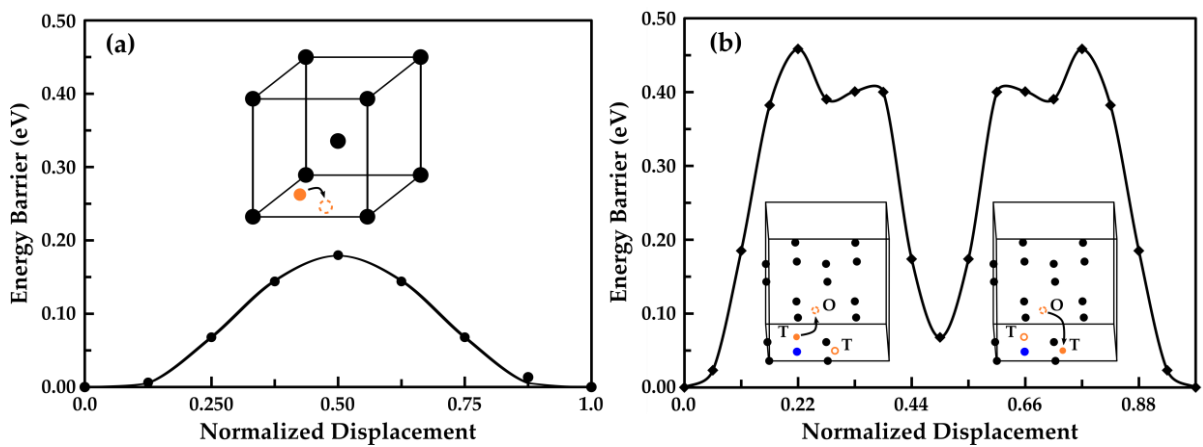

**Figure 4:** Activation energy barrier for hydrogen diffusion from one tetrahedral site to another (a) in bulk niobium and (b) in niobium near a doped nitrogen atom. Niobium, hydrogen and nitrogen atoms are represented by black, orange and blue spheres, respectively.

*3.3 Effect of N doping on hydride stability in Nb*

In addition to the thermodynamic and kinetic stability of hydrogen, nitrogen also affects the stability of hydride precipitates in niobium. Therefore, studying the effect of nitrogen on the stability of beta niobium hydride ($\beta - $NbH) precipitates and why the low hydride concentration in nitrogen treated cavity surfaces will provide understanding that can lead to increased niobium cavity performance. To accomplish this, a $\beta - $NbH unit cell (Figure 5a) was introduced in 3 x 3 x 3 periodic replication of the niobium unit cell rotated along [110] and [1$\bar{1}$0] direction (108 atoms) (Figure 5b). Next, the niobium matrix embedded with a $\beta$ − NbH precipitate was energetically relaxed to obtain an equilibrium structure followed by introducing nitrogen at different interstitial sites in niobium. The formation energy ($\Delta$E) of nitrogen at different interstitial sites in niobium embedded with a $\beta$ − NbH precipitate were calculated according to the following equation:

$$\Delta E = E_{near} - E_{away} \qquad (3)$$

where $E_{near}$ and $E_{away}$ represent the total energy of the niobium matrix embedded with $\beta$ − NbH precipitate and doped with a nitrogen atom at different octahedral sites near and away from the hydrogen atoms, respectively. Thus, a positive value of $\Delta$E represents a decrease in the energetic stability of hydride precipitates in the niobium matrix when doped with nitrogen atom. The positive formation energies in Table 1 suggest that nitrogen decreased the stability of $\beta$ − NbH precipitate in the niobium matrix (Table 1).

**Table 1:** The formation energy of niobium matrix embedded with $\beta$ − NbH precipitate and doped with nitrogen atom at different interstitial sites with respect to the nitrogen atom doped far away from the hydrogen atoms.

| Position of N | $\Delta$E (eV) |
| --- | --- |
| Octa 1 | 0.93 |
| Octa 2 | 0.78 |
| Octa 3 | 1.45 |
| Octa 4 | 1.19 |

The decrease in the stability of β − NbH with the presence of nitrogen in niobium can be understood through the electronic DOS and valence charge transfer calculations. As discussed earlier, the overlap between the states of niobium and hydrogen corresponds to the covalent bonding between niobium and hydrogen atoms. Here, the overlap corresponds to the covalent bonding in β − NbH precipitate (Figure 5c). However, the -7 eV peak disappeared upon addition of nitrogen indicating that the states of hydrogen no longer overlap with the states of niobium (Figure 5d). Hence, the covalent bonding between niobium and hydrogen atoms was disrupted by nitrogen. Valence charge transfer calculations further explain the instability of the hydride precipitate in the presence of nitrogen in the niobium matrix. In the absence of nitrogen, charge depleted (cyan iso-surface) from the niobium atoms in the hydride precipitate was accumulated (yellow iso-surface) around the hydrogen atoms (Figure 5e). However, in the presence of nitrogen, the valence charge from niobium atoms in the precipitate and the matrix is now attracted towards nitrogen due to its partially anionic nature in niobium (Figure 5f). As a result, the niobium hydrogen covalent bonding in hydride precipitate became weaker in the presence of nitrogen, leading to a decrease in the energetic stability of β − NbH precipitates in the niobium matrix.

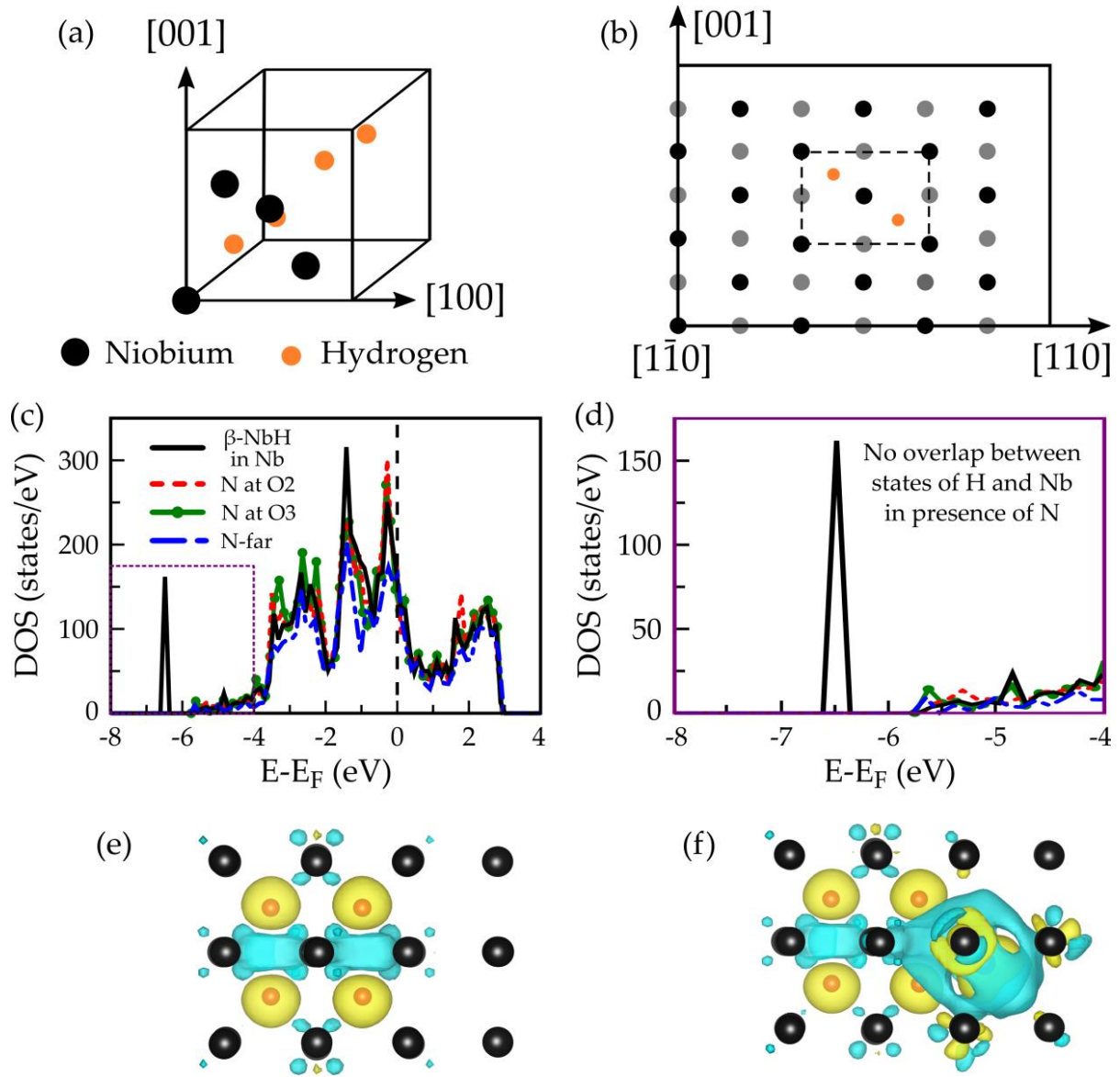

**Figure 5:** Schematic of a) $\beta - NbH$ unit cell with face centered orthorhombic structure and b) a $\beta - NbH$ unit cell embedded in the niobium matrix. Black and grey spheres represent niobium atoms on two different planes. c) Density of states for the niobium matrix embedded with $\beta - NbH$ and doped with nitrogen atoms at different octahedral (O2, O3 and far away) sites. d) Zoomed in DOS curves showing the absence of hydride peak upon addition of nitrogen in niobium. Valence charge transfer contours for e) niobium with $\beta - NbH$ and f) niobium with $\beta - NbH$ and doped with nitrogen at the far away octahedral site. Yellow and cyan iso-charge surfaces respectively represent 0.027e/Å³ of charge accumulation and depletion.

Niobium atoms are represented by dark grey spheres, hydrogen atoms are represented by orange spheres and nitrogen atom is represented by a blue sphere (below the iso-surface) in 5e and 5f.

## 4. Conclusions

In conclusion, these first-principles calculations enable identification of the mechanisms and driving forces associated with the observed low hydride concentration near the nitrogen treated niobium coupon surfaces. Nitrogen doping significantly affects the kinetic stability of hydrogen and the thermodynamic stability of hydride precipitates in niobium. In the presence of nitrogen, hydrogen binding in niobium is suppressed but the suppressing effect was dominant only in the proximity of nitrogen, approximately up to 2.5 Å away. Furthermore, the suppressing effect of nitrogen on hydrogen binding in niobium was due to the weakening of covalent bonds between the niobium and hydrogen atom, understood through electronic DOS and valence charge transfer calculations. Nitrogen also altered the minimum energy path for hydrogen diffusion from one tetrahedral site to another, leading to a notable increase in the activation energy barrier for hydrogen diffusion in the niobium lattice. Furthermore, the $\beta - $ NbH precipitate became energetically unstable near a nitrogen atom in the niobium matrix since the niobium hydrogen covalent bonding is disrupted by nitrogen. Thus, introduction of nitrogen into the surface of niobium cavities significantly decreased the thermodynamic and kinetic stability of hydrogen in niobium and consequently reduced the likelihood of hydride precipitation, which were demonstrated experimentally using simple metallographic techniques. The hydride suppression due to the presence of nitrogen could be one of the contributing reasons for the increased quality factors beyond the traditional SRF Nb recipes in cavities using this nitrogen doping recipe that are currently being produced for the LCLS-II upgrade [20,22]. However, increase in quality factors in nitrogen-doped cavities has also revealed an issue of flux trapping due to the presence of defects in bulk Nb. Further studies to understand the effect of nitrogen on the residual surface resistance and the magnetic flux trapping in niobium will help determine the overall effectiveness of nitrogen in improving the performance of niobium for SRF applications.

**Acknowledgement**

This work was supported by the U.S. Department of Energy (Award Numbers DE-SC0009962 and DE-SC0009960). A portion of this work was performed at the National High Magnetic Field Laboratory, which is supported by National Science Foundation Cooperative Agreement No. DMR-1157490 (-2017) DMR-1644779 (2018-) and the State of Florida. We thank Pashupati Dhakal from Jefferson Lab, for help with cavity based heat treatments and discussions. We are also thankful to Lance Cooley from the National High Magnetic Field Laboratory, Florida State University for many helpful suggestions and discussions. Additionally, P.G., I.A., and K.N.S acknowledge Research Computing at Arizona State University for providing HPC resources that have contributed to the research results reported within this paper.

**Supplemental material**

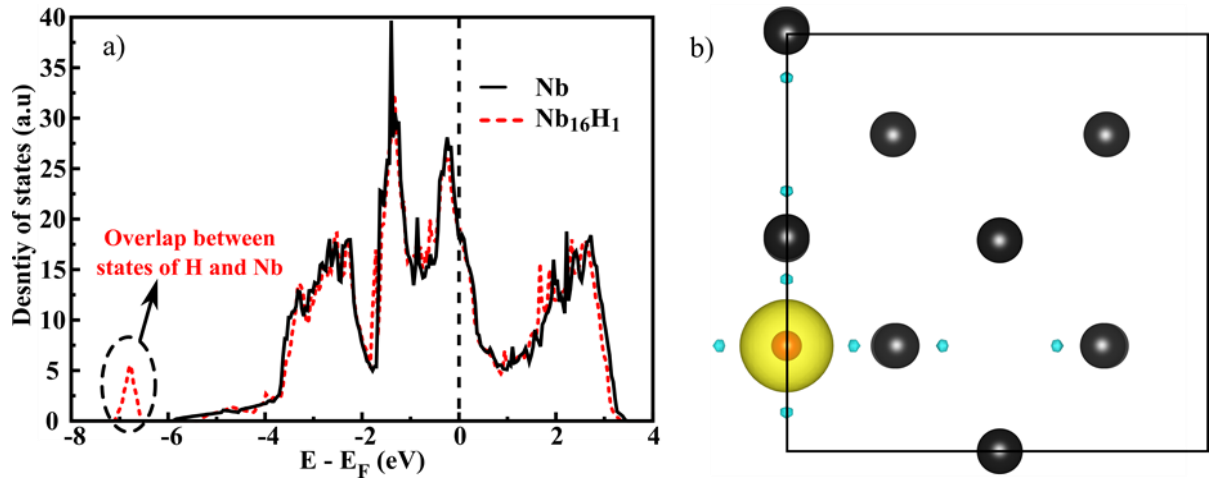

**Fig S1:** (a) Total density of state patterns and (b) valence charge density transfer contours for $Nb_{16}H_1$ (hydrogen atom at a tetrahedral site). Yellow and cyan iso-charge surfaces respectively represent an accumulation and depletion of charge of $0.027 e/Å^3$ in (b).